\documentclass[prb,aps,twocolumn,groupedaddress,superscriptaddress]{revtex4-1}
\usepackage{graphicx}  % needed for figures 
\usepackage{epsfig}    % needed for figures 
\usepackage{dcolumn}   % needed for some tables 
\usepackage{bm}        % for math 
\usepackage{amssymb}   % for math
\usepackage{amsmath}   % for math
\usepackage{mathtools} % for math
\usepackage{natbib}    % for bib
\usepackage{lipsum}
\setcitestyle{super}
\usepackage{xcolor}
\usepackage{float}
\begin{document}

\title{Distinguishing Erbium Dopants in Y$_2$O$_3$ by Site Symmetry: \textit{ ab initio} Theory of Two Spin-photon Interfaces}

\author{Churna Bhandari}
\thanks{These two authors contributed equally.}
\affiliation{Ames National Laboratory, Iowa State University, Ames, IA 50011, USA}

\author{C\"uneyt \c{S}ahin} 
\thanks{These two authors contributed equally.}
\affiliation{UNAM — National Nanotechnology Research Center and Institute of Materials Science and Nanotechnology, Bilkent University, Ankara, Turkey}
\affiliation{Department of Physics and Astronomy, University of Iowa, Iowa City, IA 52242}

\author{Durga Paudyal}
\affiliation{Ames National Laboratory, Iowa State University, Ames, IA 50011, USA}
\affiliation{Department of Electrical and Computer Engineering, Iowa State University, Ames, Iowa 50011, USA}

\author{Michael E. Flatt\'e}
\affiliation{Department of Physics and Astronomy, University of Iowa, Iowa City, IA 52242}
\affiliation{Department of Applied Physics, Eindhoven University of Technology, Eindhoven, The Netherlands}

\begin{abstract}
We present a first-principles study of defect formation and electronic structure of erbium (Er)-doped yttria (Y$_2$O$_3$). This is an emerging material for spin-photon interfaces in quantum information science due to the  narrow linewidth optical emission from Er dopants at standard telecommunication wavelengths and their potential for quantum memories {\color{black}and transducers}. We calculate formation energies of neutral, negatively, and positively charged Er dopants and find the charge neutral
%and negatively charged 
configuration to be the most stable,  consistent with  experiment. Of the two substitutional sites  of Er for Y, the $C_2$ {\color{black}(more relevant for quantum memories)} and $C_{3i}$ {\color{black}(more relevant for quantum transduction)}, we identify the former  as possessing the lowest formation energy. The electronic properties are calculated using the Perdew–Burke-Ernzerhof (PBE) functional along with the Hubbard $U$ parameter and spin-orbit coupling (SOC), which yields a $\sim$ 6 $\mu_B$ orbital  and a $\sim$ 3 $\mu_B$ spin magnetic moment, and 11 electrons in the Er $4f$ shell, confirming the formation of charge-neutral Er$^{3+}$. This standard density functional theory (DFT) approach  underestimates the band gap of the host and lacks a first-principles justification for $U$.
%which along with occupied 4$f$ split states to some degree are improved in the single particle level by the inclusion of onsite Coulomb and SOC. } 
%Here, the splitting of 4f$^3$ configuration by onsite Coulomb and SOC interactions}. 
%three unoccupied Er-$4f$-states in the middle of the band gap, confirming the formation of Er$^{3+}$. 
To  overcome these issues we performed screened hybrid functional (HSE) calculations, including a negative $U$ for the $4f$ orbitals,  with  mixing ($\alpha$) and screening  ($w$) parameters. These produced robust electronic features with slight modifications in the band gap and the $4f$ splittings depending on the choice of tuning parameters. We also computed the many-particle electronic excitation energies and compared them with  experimental values from photoluminescence.

%To Do:\\
%1. check experiments about the difference\\
%2.which journal depends on E-diff of C2 C3  possibilities: 
%    a) mat for quantum technology 
%    b) apl 
%    c) avs quantum science\\

%Arguments:\\
%1. Durga and MEF: in general in lower temperature lower symmetry site is more favorable\\
%2.Durga: 4f in core in QE, results in more reliable energies (Durga has the citation)\\
%3.MEF: setting up fermi level for low Temp such that only C2 site is the only one that is occupied\\
%4.MEF: in other applications you may want to have C3 site occupied only due to g-factor differences and anisotropy differences\\
%5. does magnetic field/strain affect the site preference? \\
\end{abstract}

\maketitle

\section{Introduction}

In general, quantum information is processed, stored,
and transmitted by employing the superposition of states of
photons or matter\cite{Goldner2015}. Rare earth-hosted crystals
are emerging as promising materials for quantum information science because at low temperatures, they exhibit
sharp optical transitions and long optical and spin coherence times. The narrow transitions allow these materials to be used as quantum light-matter interfaces (often spin-photon interfaces) or to optically control the underlying quantum states\cite{Goldner2015}. %The coherent properties of optical transitions and extended coherence lifetimes
%allow these materials to have applications in quantum
%memories and quantum computing. 
Several examples exist of wide-band-gap oxides which, when doped with  rare-earth impurities, exhibit narrow optical transitions with long coherence times and high quantum efficiency due to their partially filled $4f$ shell\cite{Jacquier2015}. These characteristics allow the fabrication and usage of these materials in highly efficient optical amplifiers, high-power lasers, and quantum information processors\cite{DiVincenzoPhysik200, DuttScience07, Yin2013, KornherPRL20}. The rare earths are not limited to  laser applications; these are excellent candidates for solid-state platforms for quantum engineering, including the development of quantum networks \cite{Hensen2015}, optical quantum memories \cite{Tittel2010}, photon sources \cite{Zhong2018}, and provide excellent hardware for quantum storage of photons\cite{Siyushev2014}. %Long-lived spin coherence times and stable optical transitions play key roles in the development of quantum engineering applications. 
The large band gap (5.6 eV) yttria (Y$_2$O$_3$), isostructural to Er$_2$O$_3$, is a very promising host for rare earth dopants due to its high chemical durability, thermal stability, and other uses for optical applications\cite{WangAdvMater05,VetroneJAP04,MuenchausenJL07,DasJPC08,Zhong2018,MaoAdvMater09}. In particular, Er is appealing as its $4f^{11}$ electron-shell configuration Er$^{3+}$ emits at the standard telecommunication wavelength %emission 
of 1.54 $\mu$m\cite{Dammakpssb03}.

Rare earth-doped yttria has been experimentally investigated for many years \cite{ManishAPL020, ScarafagioJPC019}, however many fundamental features, \textit{e.g.},  formation energies, Er-site preference, defect levels, and the band gap, are not well understood due to the absence of theoretical studies apart from a few model calculations\cite{GruberJCP85, Dammakpssb03, KlintenbergJAC98}. Therefore, a detailed first-principles study is desirable to explore the underlying physics and  electronic structure, especially due to its potential for  quantum information science.  {\color{black} Er-based quantum memories typically rely on dopant sites with inversion asymmetry; the static dipole moment allows the optical transition energy to be varied with electric field, moving the transition into and out of resonance with the hosting optical cavity. Thus a material such as Y$_2$SiO$_5$ is often chosen for quantum memories. Er-based quantum transducers, however, typically prefer sites with inversion symmetry so that the ensemble linewidth is narrower and less susceptible to local inhomogeneous electric fields. YVO$_4$, for example, is a choice for quantum transduction with inversion-symmetry Er sites.  Y$_2$O$_3$ has sites of both types: $C_{2}$, which are inversion asymmetry, and $C_{3i}$, which are inversion symmetric. Thus an important question is which of these is most stable, and whether it is possible to tune the Fermi energy, e.g. by co-doping or gate voltages, to select one or the other. The implications for spin-photon interfaces are direct: it might be that Er-doped Y$_2$O$_3$ is preferable for quantum memories or quantum transducers but not both.}

Here we study the crystal structure, defect formation energy, and electronic properties of erbium-doped yttria using improved density functional theory (DFT) methods and obtain several key results. First, from first-principles calculations, we  show that the lower symmetry site (lacking inversion, {\color{black}more suitable for quantum memories}), $C_{2}$, is energetically favorable for Er doping over the inversion-symmetric $C_{3i}$ site {\color{black}(more suitable for quantum transduction)}. Second, we find charge neutral Er$^{3+}$ doped in yttria has the lowest formation energy confirming its stability consistent with  experimental findings. Third, we investigate the electronic band structure and provide a detailed description of the Er $4f$ levels, laying a foundation critically needed for any future study of the many-body electronic excitation energies relevant for quantum information processing. Finally, an atomistic model for strongly coupled $L$ and $S$, along with a first-principles computed spin-orbit parameter, yields first and  second electronic excitation energies for the Er dopant that are in good agreement with experiment.

\section{Crystal structure and methodology}

%: Fig 1 crystal structure
\begin{figure}
    \centering
    \includegraphics[width=.49\linewidth]{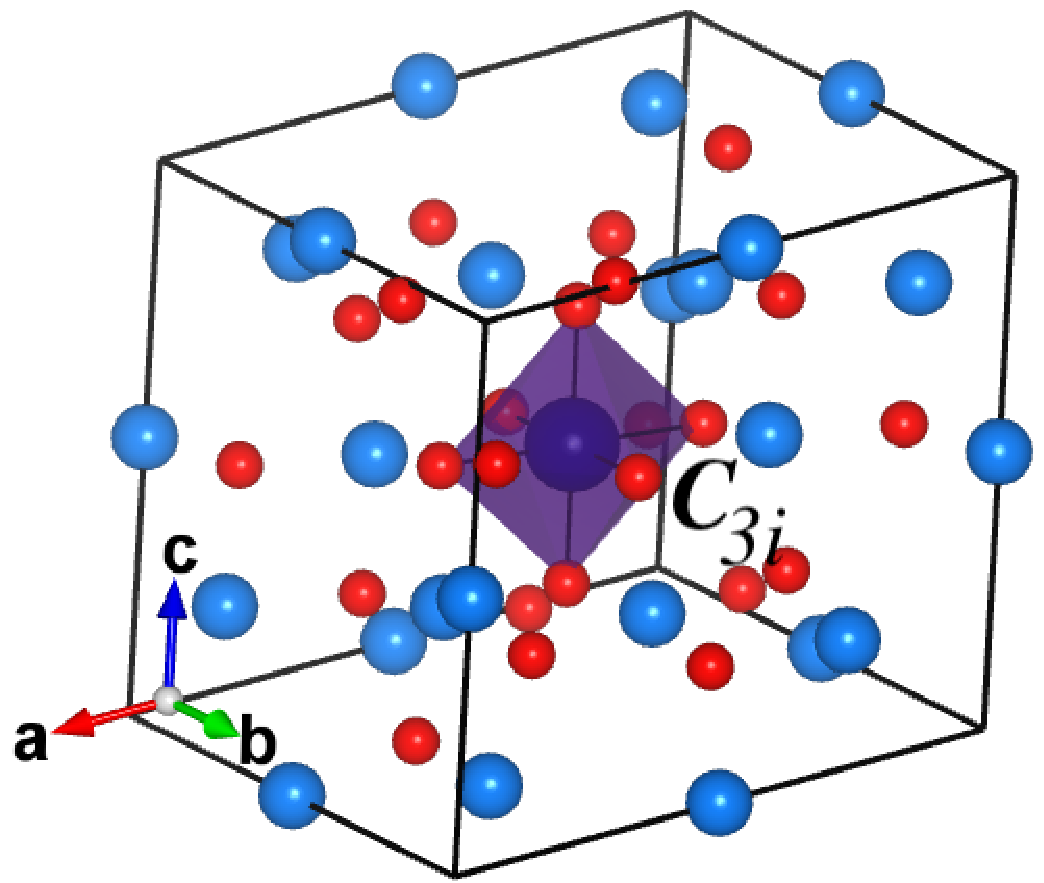}%\includegraphics[width=.49\linewidth]{C3}
    \includegraphics[width=.49\linewidth]{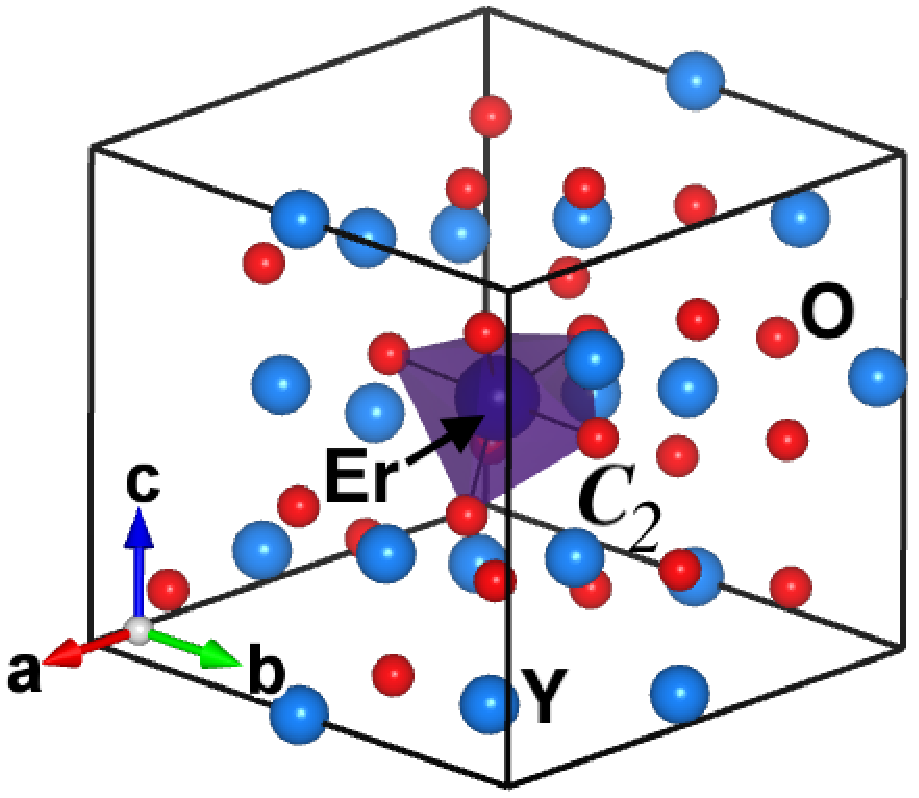}
    \caption{Crystal structure of Er-doped Y$_2$O$_3$ at $C_{3i}$({\it left}) and $C_{2}$ ({\it right}) sites and polyhedra.  Neglecting very small ($\sim 1$\%) bond length differences distorting the $C_{2}$ site, the  polyhedra show six nearest neighbour oxygens bonded to each Er. In the $C_{3i}$ site, all six nearest neighbor oxygens are equidistant from Er, whereas in the $C_{2}$ site they split into three subgroups each containing two oxygens. Only the Er at the $C_{3i}$ site has inversion symmetry.}
    \label{fig:c2c3}
\end{figure}

The primitive cell of the yttria has 40 atoms (space group Ia3 \# 206)\cite{PORTNOIChem72}. For convenience, we use a conventional bixbyite supercell with 80 atoms for defect calculations. The primitive cell consists of two types of inequivalent Y sites with $C_2 (12d)$  and  $C_{3i}$ ($4b$) symmetry in a 3:1 ratio; the oxygen is at the $24e$ site.  
%Experimentally, several Er-doped nano-crystals are reported in Ref~[\onlinecite{MaoAdvMater09}] in the range Er- $1-5$\%. 
To simulate the Er dopant one of the thirty-two Y atoms in the  Y$_2$O$_3$ cell is substituted with Er (corresponding to  3.125\%~Er doping).  This is the lowest Er dopant concentration that we were able to simulate and it corresponds to the experimentally identified optimal 1-5\% Er doping for a photoluminescence study\cite{MaoAdvMater09}.
 Er at a $C_{3i}$ site results in a structure symmetry R$\Bar{3}$ with space group  no. 148, whereas Er at a $C_{2}$ site reduces to a very low symmetry structure $C_2$ with space group no. 5 as shown in Fig. \ref{fig:c2c3}.

Two independent {\it ab initio} computing methods, the Vienna Simulation Package (VASP)\cite{vasp1} and Quantum Espresso (QE), were used to solve the Kohn Sham eigenvalue equations 
with Perdew-Burke-Ernzerhof (PBE) exchange and correlation functionals. PAW pseudopotential with full 4$f$ electrons and ultrasoft pseudopotential with frozen $4f$ states were used in the VASP and the QE calculations, respectively.  The structures were relaxed independently.  Two methods were used to provide additional support for the  identified very small formation energy difference for Er at the C$_2$ site versus the  C$_{3i}$ site.  The distinct formation energies appear to emerge from small bond length distortions associated with the C$_2$ site. 

Neglecting these small distortions,  Er atoms occupying either of the two inequivalent Y-sites have six nearest neighbour oxygen atoms. Er at the C$_{3i}$ site has inversion symmetry and six coordinated oxygen atoms with equal bond lengths of 2.29~\AA. However, Er at the C$_2$ site lacks inversion symmetry and the six oxygens have varying bond lengths with essentially the same average as that of the C$_{3i}$ site : two with 2.25\AA, two with 2.28\AA, and two with 2.34\AA. %The Er with anisotropic or distorted coordinates with oxygens and with lower symmetry structure may explain why C$_2$ is preferred over C$_{3i}$. 
{\color{black}We  note that the different symmetry of these sites has important implications for the distinct applications of these dopants for spin-photon interfaces. The C$_{3i}$ site has inversion symmetry and thus is nearly insensitive to electric field noise; this makes the C$_{3i}$ site Er more useful for quantum transduction (\textit{i.e.} microwave to optical photon transduction). Meanwhile the C$_2$ site is sensitive to electric fields, however this can also be harnessed to achieve individual addressability through a transition energy shift with electric field, which is useful for quantum memory applications. Studies of the coupling to electric fields, or photons, are  beyond the scope of this paper.}

In the VASP calculations, we used Dudarev's\cite{DudarevPRB98} on-site electron-correlation for Er $4f$ states, and the spin-orbit interaction \cite{KressePRB00} within the augmented plane wave (PAW) scheme\cite{vasp2, KressePRB99}.  Using the noncollinear spin orbit calculations including on-site electron correlation (PBE$+U+$SOC), both spin and orbital magnetic moments are calculated.
A total energy cut-off of 500 eV for the plane-wave basis set expansion was used to achieve the self-consistent charge density and the total energy. 
In QE \cite{quantumespresso,quantumespresso2}, following convergence tests, we chose a uniform $2\times2\times2$ Monkhorst-Pack {\bf k}-grid for reciprocal space integration and a plane wave energy cut-off of 90 Ry. The system was relaxed by a quasi-Newton algorithm until an energy convergence threshold of 10$^{-6}$ Ry and a force convergence threshold of 10$^{-4}$ Ry/Bohr were reached for self consistency. Similar convergence parameters were used in the VASP calculations.

We  primarily used the HSE06 functional to  study the electronic properties. 
%Here, we study the crystal structure, defect formation energy, and electronic properties of erbium-doped yttria using density functional theory (DFT) and obtain the following key results. Neutral Er-doped yttria is stable relative to other possible charged defects. Er favors the $C_2$ over the $C_{3i}$ site at low temperatures. The Er manifests the Er$^{3+}$ valence state accompanied by the emergence of three unoccupied $4f$-states within the gap in PBE + $U$ calculations. 
The Heyd–Scuseria–Ernzerhof (HSE) screened hybrid functional\cite{EdwardJCP08} is
\begin{eqnarray}
    E_{xc}^{HSE}&=&\alpha E_{x}^{HF,SR}(w) + (1-\alpha)E_x^{PBE,SR}(w)\nonumber \\
   & & +E_x^{PBE,LR}(w)+E_c^{PBE},
\end{eqnarray}
where $E_{x}^{HF,SR}$ is the short-range Hartree-Fock (HF) exchange,  $E_x^{PBE,LR}$ ($E_x^{PBE,SR}$) the long-range (short-range) part of the PBE exchange functional, and $E_c^{PBE}$ is the PBE correlation functional. The standard HSE06 approximation\cite{HSEJCP06, JalliIEE06} uses HF with the mixing parameter $\alpha=0.25$ and the
screening parameter $w=0.2$ \AA$^{-1}$. %{\color {blue} However, this approach, in Er doped yttria, pushes unoccupied $4f$-states above the band gap, producing resonant states. 
%incompatible with the experimentally-observed narrow linewidths of optical transitions. 
We further investigate the $4f$ level splittings and band gap by tuning the HSE parameters, including the screening parameter and the  value of the Hubbard $U$ for the Er $4f$ states. We use PBE for initial convergence and then refine with the HSE form. 
%Calculations with a finite but negative value of Hubbard $U$ on in addition to the HSE06 improve the key band features over calculations with PBE + $U$ or with  HSE06, including the alignment of the {\color{blue} occupied} and unoccupied $4f$-states. 
%For instance, the occupied $4f$ split states become more localized along with the positioning of the spin down 4$f$ much closer to the Fermi level. The splittings of occupied and unoccupied $4f-4f$ and $4f-5d$-states are significantly improved. %relative to experimental information.
%{\color{red} ------------\\
%I am not sure about the following statements without any reference\\
%------------\\}
%${\color{blue} We note that the sharp atom-like transitions of Er$^{3+}$ are transitions within 4$f^{11}$ configuration. Such transitions are well described by many-electron levels of 4$f^{11}$, which are split due to the combination of the 4$f$-4$f$ Coulomb interactions and the 4$f$-spin-orbit interaction, with further finestructures due to crystal-field interaction.}

%Specifically, QE is used to compute the formation energy with a frozen $4f$-core %\cite{quantumespresso,quantumespresso2}, and the results are confirmed with the VASP calculations. 

%Ultra-soft pseudopotentials describe the core electrons of the host and dopant atoms. 

Usually dopants with similar ionic radii to the atoms for which they substitute do not significantly distort the crystal structure of host materials at low doping concentrations. We notice that relaxing Er-doped yttria introduces minimum distortions in the lattice constant and atomic positions due to the similar atomic radii of Er  and Y ions. The relaxed lattice constant of the pure yttria supercell is 10.62~\AA, which is in excellent agreement with  experimental measurements \cite{Hanic1984} (10.60~\AA).  This is a considerable improvement compared to previous studies\cite{Badehian2013}. Inclusion of the $4f$ electrons yields similar lattice constants for pure and doped systems within the VASP PBE calculations. We also relaxed the structure using HSE06, which yielded 10.58~\AA, $\sim 1\%$ smaller than the experimental lattice constant of yttria, without any significant change in the electronic structure properties.
%Including the full component of $4f$-electrons yields similar lattice constants for pure yttria as in QE. For Er-doped structures, we find the optimized lattice constant to be $10.65$~\AA~, expanded by $\sim 0.4\%$ more than in pure yttria, which is expected in the GGA calculation. 

\section{Formation Energy}
When Er is substituted in yttria, it can replace a yttrium atom at one of the two symmetrically nonequivalent sites. 
Since  Y has a ${3+}$ electronic configuration, and photoluminescence emission energies associated with Er$^{\rm 3+}$ have been observed in doped material, we expect the neutral supercell configuration with Er in the $3+$ state to be stable at least under some conditions.  We compute the formation energy of the Er ion in  yttria to verify the stable state. 
The formation energy of different charged configurations of the Er dopant is calculated from \cite{Zhang1991}
\begin{align}
    E_{f}^q(\epsilon_F)=E^q_{\text{tot}}-E^{\text{bulk}}_{\text{tot}}&+E_{\text{corr}}+ \sum_i n_i \mu_i\\&
   \nonumber +q(E_{\text{VBM}}+\epsilon_F+\Delta_{q/b} ).
\end{align}
Here the first two terms stand for the total energy differences between the pristine bulk system and  supercell with a defect. $E_{\text{corr}}$ is the correction term arising from the interaction between charged supercells. $\mu_i$ is the chemical potentials of the added (Er with negative n) and subtracted (Y with positive n) atoms. $\epsilon_F$ and $E_{\text{VBM}}$ are the Fermi level and the valence band maximum of the pristine supercell. $\Delta_{q/b}$ represents the potential alignment term between the charged cell and the valence band gap of the bulk supercell.

To obtain the correction term ($E_{\text{corr}}$), which eliminates artifacts that arise in \textit{ab initio} calculations with charged supercells due to the periodic boundary conditions, we have used three different methods and modified the final energies accordingly. The first method is the first-order Makov-Payne correction  \cite{Makov1995}. The correction term is $q^2\alpha/{2\epsilon L}$ where $q$, $\epsilon$, $\alpha$, and $L$ represent the additional charge, the dielectric constant of the host material, the  lattice-dependent Madelung constant ($\alpha=2.8373$ for a simple cubic supercell), and  the linear size of the supercell, respectively. This correction can be amended to get $E_{corr}$ in the units of eV, such that $E_{MP}=14.39952 \alpha q^2/2 \epsilon L$ with the same definitions above, and L is in the units of \AA. We use the experimental static dielectric constant of 18.1. With the relaxed supercell size mentioned above, the first-order Makov-Payne corrections are 106.284 meV for +1, one hole or -1, one electron, and  425.138 meV for +2,  two holes or -2, two electrons, charged supercells. 

As a second option, we used the QE software to directly calculate the Makov-Payne correction to the total energy and obtain the same numerical results. In a third approach we employed a completely different correction scheme, the method proposed by Freysoldt, Neugebauer, and Van de Walle (FNV) \cite{Freysoldt2009,Freysoldt2011},  which is modeled by the sxdefectalign software. Similar to the previous calculations, we get a charge correction of about 103 meV and 414 meV for a single and double-charged supercell, respectively. We also utilized the sxdefectalign software \cite{Freysoldt2009,Freysoldt2011} to obtain the potential alignment, $\Delta_{q/b}$, which is usually the smallest correction term in the formation energy calculations.

\begin{figure}[htb!]
 %   \centering
    \includegraphics[width=.99\linewidth]{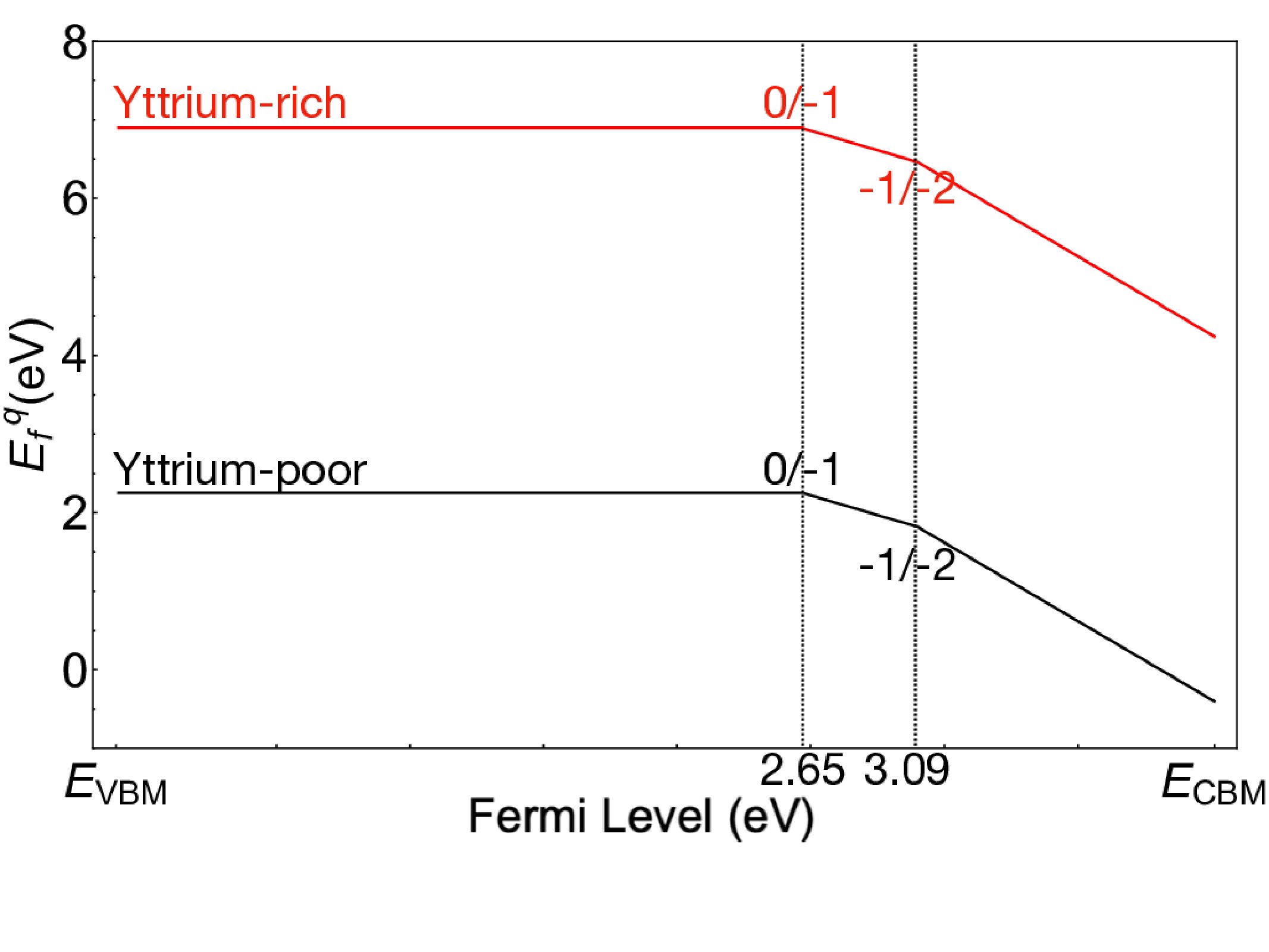}
    \caption{Formation energies of neutral and negatively charged Er ions in Y$_2$O$_3$ host as a function of the Fermi level. We select two growth conditions: yttrium-rich (red lines) and yttrium-poor (black lines) by changing the chemical potential of the Y atom in the formation energy.}
    \label{fig:formation}
\end{figure}

We have computed formation energies for the $C_{3i}$ site for 5 differently charged configurations: -2 (Er$^{+1}$), -1 (Er$^{+2}$), 0 (Er$^{+3}$) +1 (Er$^{+4}$), +2 (Er$^{+5}$) of the erbium-doped yttria  and include only the lowest energy -2, -1, and 0 (the neutral) charged states  in  Fig.~\ref{fig:formation}. %%{\color{red} These charged configurations are in the whole supercell, not specifically on the Er atom}. 
As such, the formation energies of the positively charged Er ions are always much higher for all the Fermi levels. We find that the neutral impurity, which corresponds to the Er$^{3+}$ state, is the most probable one for a Fermi level that is between the valence band maximum (0 eV) and 2.65 eV.The charge transition level (from 0 to -1 (Er$^{2+}$) shown as 0/-1 in the  Fig.~\ref{fig:formation})  appears at 2.65~eV, and another charge transition from -1 to -2 (Er$^{+}$) occurs at a Fermi level of 3.09 eV with respect to the top of the valence band. We also calculated formation energies with the yttrium-poor (oxygen-rich) and yttrium-rich (oxygen-poor) conditions. Yttrium-poor conditions yield to lower formation energy as expected giving Er impurities a higher chance to substitute for Y atoms.  For  higher doping levels up to the energetic position of the bottom of the conduction band, a -2 charge state is the most probable configuration, however, we are unaware that erbium has been observed in yttria in these charged states. It is possible that for current materials the Fermi level is always near the valence band for  bulk ungated yttria, for which charge-neutral erbium $+3$ has the lowest formation energy. However, yttria is an excellent dielectric material
\cite{Gurvitch1987} in which the Fermi level in a sufficiently thin material should be tunable by adjusting a gate voltage. 
Additionally, we also note that we do not find any qualitative difference in the formation energies for Er substituted in the $C_2$ site with a slight shift in the energies to 2.64 eV and 3.05 eV for the 0/-1 and -1/-2 charge transitions, respectively.

In addition, the lower formation energy  with respect to YO$_2$, ErO$_2$, and Er$_2$O$_3$ confirm the thermodynamical stability of Er doped Y$_2$O$_3$ (\textit{e.g.}, ErY$_{31}$O$_{48}$). The stability conditions and the enthalpy for the host (yttria) are:
\begin{align}
    2\mu_Y^{Y_2O_3}+3\mu_O^{Y_2O_3} =E_{Y_2 O_3}\\
    \Delta H_f =E_{Y_2O_3} -2\mu_Y^0 -3\mu_O^0
\end{align}
where $\Delta H_f$ and $E_{Y_2O_3}$ are the heat of enthalpy and the chemical potential of Y$_2$O$_3$, $\mu_Y^{Y_2O_3}$ and $\mu_O^{Y_2O_3}$ are the chemical potentials for yttrium and oxygen within the yttria, and $\mu_Y^0$ and $\mu_O^0$ are the chemical potentials for the yttrium atom and oxygen atom in bulk, respectively. We calculated the chemical potentials of a  yttrium atom and an oxygen atom from the bulk four-atom conventional Y crystal and an oxygen molecule. The enthalpy of formation is -1.346 Ry (-18.313 eV) (experimentally -19.7 eV \cite{Robie1978}), a negative number indicating that yttria is stable against phase separation into bulk Y and gaseous O$_2$. One Er substitution out of thirty-two Y should still keep the enthalpy of formation negative at least within the 8\% miscibility limit provided by Ref.~\onlinecite{doi:10.1063/1.2214299}.   

The chemical potentials of yttrium and oxygen are bound by the calculated values such that
\begin{align}\label{Eq:thermocond}
    \frac{1}{2}\Delta H_f^{Y_2O_3} +\mu_Y^0 < \mu_Y^{Y_2O_3} < \mu_Y^0 \\
    \frac{1}{3}\Delta H_f^{Y_2O_3} +\mu_O^0 < \mu_O^{Y_2O_3} < \mu_O^0,
\end{align}
where a chemical potential closer to the heat of enthalpy ($\Delta H_f$) indicates a poor condition, such as oxygen-poor or yttrium poor, and a chemical potential closer to the bulk value ($\mu_Y^0$ and $\mu_O^0$) is an oxygen or yttrium rich growth condition. 

In order to incorporate Er we calculate the formation enthalpy of  Er$_2$O$_3$ as an example of a competing phase, as the chemical potential of the Er in yttria
is bound by the enthalpy of Er$_2$O$_3$ and the chemical potential of  pure Er ($\mu_{Er}^{Y_2O_3}<\mu_{Er}^0$ and $2\mu_{Er}^{Y_2O_3}+3\mu_O^{Y_2O_3}<\Delta H_f^{Er_2O_3}$) preventing the formation of Er$_2$O$_3$ while keeping the same oxygen chemical potential in Eq.~(\ref{Eq:thermocond}). We calculate $\Delta H_f^{Er_2O_3}$ to be -18.806 eV for an 80-atom supercell. 

Additionally, we compute the monopole correction term directly from  VASP, which gives 0.14 eV for both cases when an electron or a hole is added to the supercell, which is consistent with the estimated value from the first-order Makov-Payne corrections. Both VASP that includes 4$f$ electrons as valence electrons and QE  with a frozen 4$f$-core produce similar formation energies. 

{\color{black}
\section{Site preference and Implications for Spin-Photon Interfaces}

}

Er can occupy two nonequivalent sites, the $C_2$ and $C_{3i}$, and the geometric ratio of $C_2$ and $C_{3i}$ yttrium sites is 3:1. %However, the site preference is modified by the energetics of doped structures. 
{\color{black} There is a difference in the behavior and utility of the two sites for spin-photon interfaces. The $C_2$ site possesses a static electric dipole moment due to the lack of inversion symmetry, making it more useful for quantum memories in which single ion optical transitions are moved into and out of resonance with an optical cavity using an electric field. The $C_{3i}$ site, however, lacks that static electric dipole moment and is more useful in a narrow linewidth ensemble for quantum transduction.}  %Experimentally, the optical properties of the defect can be controlled with temperature if the energy difference between these two sites is known, and the Fermi energy can be manipulated also to lie between them, \textit{e.g.} with co-doping. 
Experimentally electronic transitions of Er$^{}$ ions from both the $C_2$ and $C_{3i}$ sites are seen in photoluminescence\cite{MaoAdvMater09,GruberJCP85, Dammakpssb03}, which suggests the energy difference between the two sites is smaller than the room temperature thermal energy.  
Both PBE + $U$ and HSE06 methods predict a lower energy for $C_2$ as the site symmetry is lower than the $C_{3i}$ site. We find a 2.6~meV lower formation energy for the $C_2$ site  compared to the $C_{3i}$ site within PBE + $U$ with a sensible choice of $U_{eff}=4$ eV. HSE06 yields a 8.3~meV (96.32~K) lower energy for the $C_2$ site. 

{\color{black} Although the preference energy difference is smaller than the room temperature thermal energy, most quantum devices are intended to operate at temperatures far below 1K due to the need for microwave cavities with several GHz resonant frequencies that are depleted of thermally excited photons. Thus the site energy difference for Er far exceeds the thermal energy at temperatures of relevance for quantum devices. The Fermi energy could be positioned between that of the $C_2$ and the $C_{3i}$ sites by co-doping with an electron acceptor; another method would be to apply a gate voltage to deplete these electrons from the material. Yttria is a known gate electrode which can sustain substantial voltages without leakage. If the material is to be used for quantum transduction it will be important to guarantee that the $C_{3i}$ sites are all neutral; this can be done by co-doping with electron donors or gate voltages as well. The presence of the $C_{2}$ sites nearby, which will remain under these conditions, is not significantly detrimental for quantum transduction. In addition, a gate voltage could  be used to shift the optical transition energies of the $C_2$ sites far from the $C_{3i}$ optical bandwidth relevant for quantum transduction. Thus it appears that with proper choices of co-doping and gate voltages yttria doped with erbium makes an excellent choice for quantum memories and for quantum transduction.}

{\color{black} The VASP calculations are performed with a smaller 40 atom yttria unit cell.}  
An independent frozen-$4f$ core approximated pseudo-potential calculations {\color{black}  with a larger 80 atom cubic cell also show similar results, confirming the independency of the cell shape and size for Er-site preference}. Our calculated energy difference suggests that site-specific Er selection is possible by adjusting the temperature and changing the Fermi energy of the material with doping or electrostatic gates.\\  %However, at higher temperatures, Er likely fluctuates between two sites which may lead to admixtures in the optical spectra.

%Interestingly, GGA shows the opposite 12.4 meV less energy for $C_{3i}$-site.
\section{Electronic structure}
\subsection{PBE and PBE + $U$}
%{\color{red}The Er is embedded in the solid state matrix of $4d$ oxides, yttria. }
%As we discuss in the later paragraph, the standard DFT severely underestimates the electronic band gap of the host as well as the splitting of $4f$-levels.

It is well-known that local and semi-local exchange functionals severely underestimate the band gap of semiconductors and insulators due to the self-interaction error\cite{HufnerAdnace94, DudarevPRB98}. Our VASP calculations with the PBE functional produce a band gap for yttria $\sim 4.2$ eV, similar to  previously reported results\cite{RamzanCMS013} with the same method and  functional. Previously reported calculations using the orthogonalized linear combination of atomic orbitals (OLCAO) and the linear muffin-tin orbital within the atomic sphere approximation (LMTO-ASA), both with the local density functional, show a slightly higher band gap of $\sim 4.5$ eV (at $\Gamma$)\cite{XuPRB097, MuellerPRB96}. All these theoretical band gaps are off by $\sim 1.6-1.9$ eV from experimental measurements (5.5 eV [Ref.~\onlinecite{TomikiJPS86}], 5.6 eV [Ref.~\onlinecite{Zhang1998}], and 5.9 eV [Ref.~\onlinecite{Ahlawat2017}]) except that earlier OLCAO calculations\cite{ChingPRL90} in the local density approximation yielded a much smaller direct band gap of 2 eV at $\Gamma$  which the authors ascribed to a unique type of bonding (Y$_2^{+2.16}$O$_3^{-1.44}$) in yttria.\cite{ChingPRL90}

%It is well-known that local and semi-local exchange functionals severely underestimate the band gap of semiconductors and insulators due to the self-interaction error\cite{HufnerAdnace94, DudarevPRB98}. { From our PBE calculations we find a band gap $\sim 4.2$ eV similar to previous local density approximation calculations of  $\sim 4.5$ eV }(at $\Gamma$)\cite{XuPRB097} using the orthogonalized linear combination of atomic orbitals (OLCAO) method  and the linear muffin-tin orbital method (LMTO) within the atomic sphere approximation (ASA) in the local density approximation (LDA)\cite{MuellerPRB96}, and VASP calculations within PBE\cite{RamzanCMS013} (4.3 eV). All these theoretical band gaps are off by $\sim 1.6-1.9$ eV from experimental measurements (5.5 eV \cite{TomikiJPS86}, 5.6 eV \cite{Zhang1998}, 5.9 eV \cite{Ahlawat2017}) except that earlier LCAO calculations\cite{ChingPRL90}  within the DFT in the LDA formalism which yielded a much smaller direct band gap of 2 eV at $\Gamma$.}

\begin{figure}[t]
    \centering
    \includegraphics[scale=0.375]{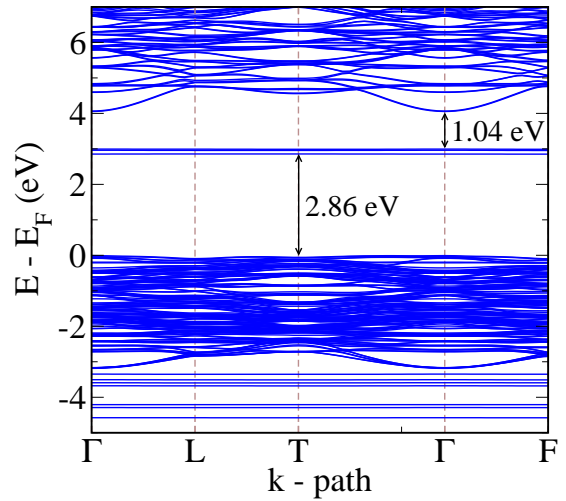}
    \caption{PBE + $U$ + SOC band structure of  Y$_2$O$_3$, with Er-substituted into the $C_{3i}$ site, along the high symmetry {\bf k-} points viz., $\Gamma=(0,0,0)$, $L= (\dfrac{1}{2}, 0,0)$, $T= (\dfrac{1}{2},-\dfrac{1}{2},\dfrac{1}{2})$, and $F=(\dfrac{1}{2},-\dfrac{1}{2},0)$ of the trigonal primitive unit cell with space group no. $148$ ($S_6 = C_{3i}$). The unoccupied $4f$ levels are split into two-manifolds, singly and doubly degenerate bands.}
    \label{fig:bandPBEU}
\end{figure}

\begin{figure} [t]
    \centering
    \includegraphics[width=0.95\linewidth]{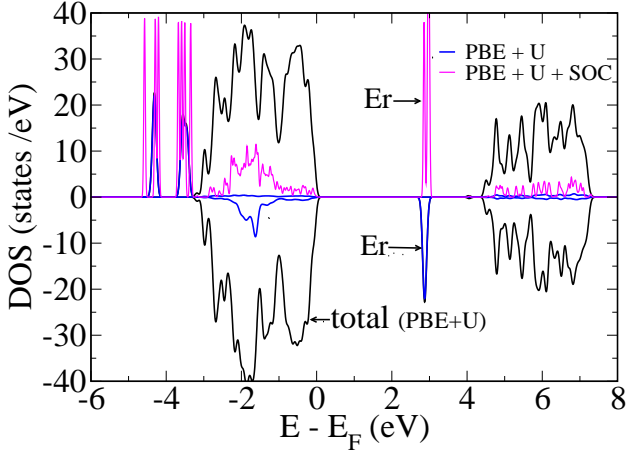}%\includegraphics[width=.45\linewidth]{C2PBE + $U$4primitive_dos.eps}
    \caption{Total density of states of Er-doped Y$_2$O$_3$ in $C_{3i}$ site computed with PBE + $U$ and partial density of states (PDOS) of Er obtained with PBE + $U$ in {\it blue} and PBE + $U$ + SOC in {\it magenta}. Positive and negative values represent DOS for spin-up and spin-down electrons in PBE + $U$.
    The unoccupied Er $4f$ states lie in the middle of the band gap, which further split into two manifold due to SOC.}
    \label{fig:dos}
\end{figure}

The PBE + $U$ + SOC approach does better  describing the band alignment as well as the splittings of strongly correlated $4f$ levels\cite{Anisimov1997}. 
%{\color{red} add reference to this in general}
The electronic band structure of Er-doped yttria at the $C_{3i}$ site (Fig. \ref{fig:bandPBEU}) is obtained with PBE + $U$ + SOC, where $U_{eff}=4$ eV for the Er $4f$ orbitals. The band structure shows  three  localized bands in the gap region (2.86 eV above the Fermi level) of the host material. The inclusion of SOC leads to additional splittings of the $4f$ levels, i.e., the threefold degenerate unoccupied  states in PBE + $U$ (see Fig. \ref{fig:dos}) split into two manifolds.

%: PBE + $U$ band structure C3 site U=4eV
For a more detailed understanding of the electronic structure, we show the electronic density of states computed with $U_{eff}=4$ eV for the $4f$ states in Fig.~\ref{fig:dos} for Er doped into the $C_{3i}$ site. The occupied Er $4f$ states split into three manifolds, with the lowest two DOS peaks substantially localized, whereas the states near the Fermi level are delocalized. The unoccupied three $4f$ levels lie 2.86 eV above the Fermi level. The Er $5d$ levels are located at 4.5 eV above the Fermi level. Our choice of $U=4$ eV is based on the energy difference between the occupied $4f$ and unoccupied $5d$ states, which is $\sim 5.6$ eV, and it agrees with  optical spectra \cite{MaoAdvMater09}. We note that there is not much difference in the DOS for Er at the $C_{3i}$ sites from Er at the $C_2$ sites (not shown here).
%{\color{red} The bare Er atom consists of  $6s^2$ and $4f^{12}$ as the valence electron orbitals, whereas after implantation in yttria the dopant becomes Er$^{3+}$ ($4f^{11}$) valence orbitals, consistent with the band structure and DOS.} 
For Er at either site we find a  spin magnetic moment of $3$~$\mu_B$ and an orbital magnetic moment of $\sim 6$~$\mu_B$ in PBE $+U+$SOC, which is consistent with an Er$^{3+}$ ($4f^{11}$) shell configuration. %{\color{blue} Although this approach produces sensible $4f$-splittings, the band gap of the host is off by about 1 eV from the experiment.}   

Now we estimate the many body excitation energies at the atomistic level\cite{mackintosh1991} employing the following relation,
\begin{equation}
    E_J = \dfrac{\lambda}{2}[J(J+1)-L(L+1)-S(S+1)],
\end{equation}
where $J= L+S$ is the total angular quantum number with $L$ the orbital and $S$ the spin quantum numbers, and $\lambda$ is a SOC parameter. We estimate the excitation energy between the ground state $^4I_{15/2}$ manifold and the excited $^4I_{13/2}$ manifold, 
\begin{equation}
\Delta E^{01} = \dfrac{2E_{\rm soc}} {(J_{max}+1)},
\end{equation}
using the first-principles value calculated for
\begin{equation}
\lambda= \dfrac{2E_{\rm soc}}{J_{max}(J_{max}+1)},
\end{equation}
where $E_{\rm soc}$ is the spin-orbit energy obtained by taking the total energy difference calculated with and without SOC, i.e., $E_{\rm soc}$= $E_{\rm total}$(PBE$+U+$SOC) $-$ $E_{\rm total}$(PBE$+U$) and $J_{max} =  15/2$. The estimated value of $\Delta E^{01}$ is 0.71 eV, which is in good agreement with the experimental value of $0.807$ eV (1535~nm wavelength)\cite{MaoAdvMater09} in photoluminescence obtained for Er-doped nanocrystals excited with 488 nm photons at room temperature. Similarly, for the ground state to the second excited state, the calculated energy is 
\begin{equation}
\Delta E^{0 2} = \dfrac{2E_{\rm soc}(2J_{max}-1)} {J_{max}(J_{max}+1)} = 1.33 {\rm eV},
\end{equation}
which is also close to the experimental value of 1.265 eV (980 nm wavelength). The discrepancy in our calculated values is reasonable due to the simplified effective atom approximation for the excitation energy.
%E_{soc}=3.04 for relaxed structure
\subsection{Hybrid functional}

\begin{figure}
    \includegraphics[scale=0.375]{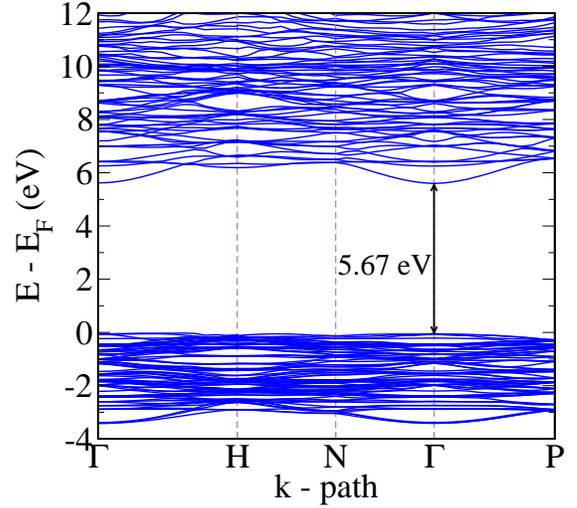}
    \caption{HSE06 band structure of Y$_2$O$_3$ along the high symmetry {\bf k}-points: $\Gamma=(0,0,0)$, $H= (\dfrac{1}{2}, -\dfrac{1}{2},\dfrac{1}{2})$, $N= (0,0,\dfrac{1}{2})$, and $P=(\dfrac{1}{4},\dfrac{1}{4},\dfrac{1}{4})$ of the primitive cell space group no. 206 (Ia3).}
    \label{fig:HSE06pristine}
\end{figure}

\begin{figure*}[!ht]
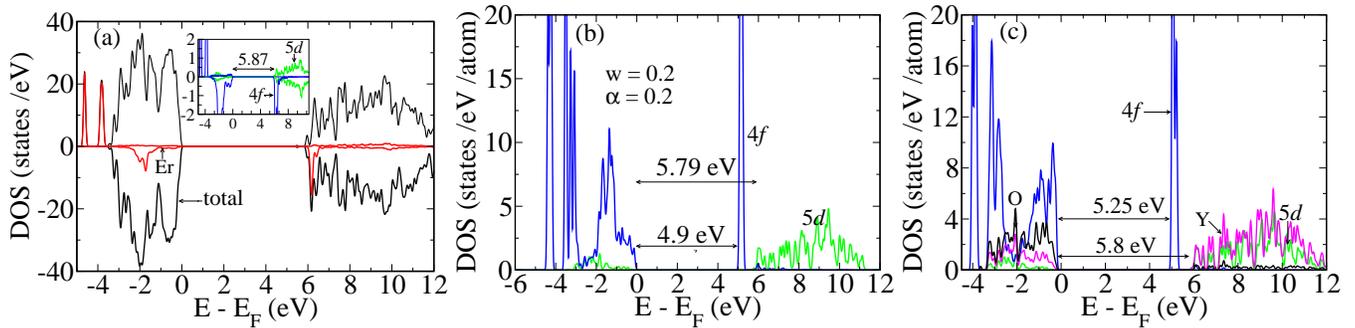

    \centering
    \includegraphics[width=.325\linewidth]{C3HSEdos.eps}
    \includegraphics[width=.325\linewidth]{HSECORdos.eps}
    \includegraphics[width=.325\linewidth]{C3HSEminusUSOCdos.eps}
    \caption{(a) Spin-polarized total DOS {\it in black} and Er PDOS {\it in red} of Er-substituted  Y$_2$O$_3$ in $C_{3i}$-site and HSE06, and (b) Er PDOS computed with HSE and SOC  with $\alpha=0.2$ $w=0.2$ and (c) HSE06,  $U$ ($U_{eff}=-2$ eV), and SOC. The positive and negative values represent the DOS for spin-up and spin-down electrons.
    The inset in (a) shows the partial density of states of Er $5d$ { in green} and $4f$ { in blue}. The band gap as well as the $4f$-$5d$ splittings slightly change depending the choice of the HSE parameters.}
    \label{fig:HSE}
\end{figure*}
As our PBE + $U$ + SOC calculations indicate that the Y $4d$ states are all unoccupied, the inclusion of $U$ does not affect the band gap of the host material. For the Er dopant the inclusion of $U$ only affects the Er $4f$ states without altering the band gap of the host material.
 Whereas PBE + $U$ produces  dopant states in the gap region, it fails to predict the correct band gap of the host material. To overcome this issue, we also performed an HSE06 band structure calculations for pristine yttria as shown in Fig. \ref{fig:HSE06pristine}. There is no unique choice of $\alpha$ and $w$ that reproduces the accurate band gap for all materials\cite{FrancescJCC017}. For instance $\alpha=0.25$ and $w\sim 0.38$ \AA$^{-1}$ is shown to reproduce the correct band of anatase TiO$_2$\cite{anatase}, whereas $\alpha=0.2$ and $w=0.2$ \AA$^{-1}$ works for rutile TiO$_2$\cite{rutileHSE}.
However, for yttria  the standard HSE06 calculation produces a direct band gap of 5.67 eV at $\Gamma$, which is in better agreement with experiment (5.6 eV) than the previously reported theoretical value \cite{RamzanCMS013} of 6 eV.

Next, we discuss our HSE results for Er-doped yttria. %As there is not much difference between the overall electronic structures with Er-doped at $C_{3i}$ and $C_2$ sites, we only show DOS for $C_{3i}$-site. 
It is of interest to examine the effect of HSE on the Er $4f$ and $5d$ levels, which has not previously been explored to our knowledge.  The only realistic comparison of our theoretical results is to the experimentally measured excitation energy between occupied $4f$ and unoccupied $5d$ states,  which is $\sim 5.8$ eV. 

To check the consistency of our approach, we present the three different results by tuning the HSE parameters viz., (i) the standard HSE06 without SOC, (ii) HSE with $\alpha=0.2$ and $w=0.2$ \AA$^{-1}$ and SOC, and (iii) HSE06 and SOC with $U_{eff}=U-J= -2$ eV for Er $4f$, as shown in Fig.~\ref{fig:HSE}.
As expected, HSE06 improves the band gap significantly (6 eV), similar to its effect in calculations for the undoped host material. 
The unoccupied $4f$ levels are mixed with Y and O conduction states, which does not occur in PBE + $U$. The excitation energy between the occupied $4f$ and unoccupied $5d$ is $\sim 5.87$ eV in agreement with experiment. 
The tuned HSE with $\alpha=0.2$ and $w=0.2$ \AA$^{-1}$ slightly reduces the band gap and $4f$ splittings. As expected, reducing the exact short-range Hartree-Fock exchange $E_{x}^{HF,SR}$ reduces the $4f$-$4f$ splittings. The $4f$-$5d$ excitation energy is now much closer to  experiment. In addition, it also leads to a slight decrease in the band gap for the host.

 The third set of calculations above, (iii), have a negative $U_{eff}= -2$ eV, which is theoretically possible when the exchange term $J$ is larger than the on-site Coulomb term\cite{MicnasRevMod90, NakamuraPhysica09, HasePRB07} $U$, especially for $sp$-like delocalized states. Now, as expected, this approach further reduces the $4f$-$4f$ splittings due to the negative value of $U_{eff}$, whereas the $4f$ to $5d$ excitation energy remains more or less the same.
Overall the HSE results obtained with  different parameters are very similar qualitatively. Further fine tuning of these parameters and their validation requires additional experimental data.
  %The computed  many-particle electronic excitation energies (not shown here) are similar to that found in PBE + $U$ + SOC.} 

%For clarity, we show LDU + SO calculated band structure in Fig. \ref{fig:bandPBE + $U$}.  

%The HSE06 improves the band gap significantly, which is excellent agreement with experiment\cite{RamzanCMS013}.
% We note that the position of  Er-$5d$ and Y/O-derived band gaps are still off from the experiment.

\section{Conclusion}
By performing DFT calculations, we investigated the dopant formation energy and the electronic structure of  Er doped yttria in order to assess the utility of the material for quantum devices such as quantum memories and quantum transducers. Our results indicate that neutral Er is lower energy in the $C_2$ site relative to the $C_{3i}$ site,  which may enable controlled preferential positioning of Er in that specific site during growth. The formation energy calculations show that the charge-neutral Er dopant is more stable than  charged Er dopants. 
 The electronic structure obtained with PBE + $U$ calculations reveals the orbital ($\sim$ 6 $\mu_B$) and  spin ($\sim$ 3 $\mu_B$) magnetic moments, and  the occupancy (11 electrons) of Er $4f$, confirming the formation of Er$^{3+}$. 
 
%The standard DFT provides incorrect occupied and unoccupied bands, which along with occupied 4$f$ split states to some degree are improved in the single particle level by the inclusion of onsite Coulomb and SOC. 
We computed the first and second electronic excitation energies using a minimal atomistic model, incorporating the first-principles calculated spin-orbit parameter, and these excitation energies agree well with photoluminescence data.
Since our $U$ only is for the $d$ and $f$ states, and the highest valence and lowest conduction states are neither $d$ nor $f$, within standard DFT the inclusion of $U$ and SOC does not change the band gap of the host. 
%The splitting of 4f$^3$ configuration is by onsite Coulomb and SOC interactions. These results are} in line with the experiment. 
A more advanced approach using a standard HSE functional produces more robust electronic band features viz., the host band gap and the alignment of the Er $4f$ and $5d$ states, which remain more or less the same even with a small  variation of HSE parameters and $U$. 
%The electronic excitation energies computed in the latter agree qualitatively with the values obtained from PBE + $U$ + SOC.}
%, although it requires more investigation in future work.

{\color{black} Even in material with Er at both $C_2$ and $C_{3i}$ sites, the site energy preference has important implications for spin-photon interfaces. The $C_{3i}$ sites could be depleted of one electron by co-doping with electron acceptors or through a gate voltage. At the temperatures relevant for quantum devices ($<1$~K) this would yield a material with only the Er in $C_2$ sites that have the optical transitions associated with Er$^{3+}$. Such a material would be preferred for the quantum memories that benefit from a static electric dipole that permits shifting the optical transition energy into or out of resonance with an optical cavity. If the erbium-doped yttria is to be used for quantum transduction then the material should be co-doped with electron donors, or a gate voltage used, to ensure all the Er $C_{3i}$ sites are neutral. Furthermore the use of a gate voltage can permit the shift of the $C_2$ site optical transitions far away from the narrow linewidth optical band associated with the $C_{3i}$ sites and thus optimizing the material for quantum transduction. We conclude that yttria is a remarkable material whose spin-photon interfaces may be optimized for performance either as quantum memories or quantum transducers.} 

%Suppl: QE band calculations.
\acknowledgments
Work done at the Ames Laboratory was conducted for the US-DOE under its contract with Iowa State University, Contract No. DE-AC02-07CH11358. MEF acknowledges support from  NSF DMREF DMR-1921877 for studies of the difference in formation energies between the $C_2$ and $C_{3i}$ sites. Atomistic simulation of the excitation energies, as well as hybrid functional calculations of the band gap and partial densities of states were supported by the U. S. Department of Energy, Office of Science, Office of Basic Energy Sciences, under Award Number DE-SC0023393. We acknowledge fruitful conversations with Tian Zhong. %The electronic structure calculations in this work are carried by C. B. in the Critical Materials Institute, an Energy Innovation Hub led by the Ames Laboratory and funded by the U. S. Department of Energy, Office of Energy Efficiency and Renewable Energy, Advanced Manufacturing Office. We acknowlgap fruitful discussions with Tian Zhong, University of Chicago.

\bibliography{library.bib}
\end{document}